\documentclass[11pt]{iopart}
\usepackage[dvips]{graphicx}
\usepackage{graphicx}
\usepackage{epsfig}
\usepackage{amssymb,amsfonts}
\usepackage{cite}
\usepackage{psfrag}
\usepackage[x11names]{xcolor}
\usepackage{subfigure}
\usepackage{wrapfig}
\usepackage{bbold}
\usepackage{a4wide}
\usepackage{amsthm}

\renewcommand\({\left(}
\renewcommand\){\right)}
\renewcommand\[{\left[}

\newcommand{\be}{\begin{equation}}
\newcommand{\ee}{\end{equation}}
\def\bea{\begin{eqnarray}}
\def\eea{\end{eqnarray}}

\newcommand{\OP}{\omega_{\rm P}}

\begin{document}
\rightline{\normalfont{\small DESY 09-132; DCPT/09/124; IPPP/09/62}}
\title{Late time CMB anisotropies constrain mini-charged particles}
\author{C Burrage$^{1}$, J Jaeckel$^{2}$, J Redondo$^{1}$ and A Ringwald$^{1}$ }
    \vspace{3mm}

    \address{$^1$ Deutsches Elektronen Synchrotron DESY,
          Notkestrasse 85, D-22607 Hamburg, Germany \\[3mm]
    $^2$ Institute for Particle Physics and Phenomenology, Durham University,
Durham, DH1 3LE, UK
}
    \vspace{3mm}

    \eads{\mailto{clare.burrage@desy.de; joerg.jaeckel@durham.ac.uk; javier.redondo@desy.de; andreas.ringwald@desy.de}}
\begin{abstract}
Observations of the temperature anisotropies induced as  light from
the CMB passes through large scale structures in the late universe are
a sensitive probe of the interactions of photons in such
environments.  In extensions of the Standard Model which give rise to
mini-charged particles, photons propagating through transverse
magnetic fields can be lost to pair production of such particles.
Such a decrement in the photon flux would occur as photons from the
CMB traverse the magnetic fields of galaxy clusters.  Therefore late
time CMB anisotropies can be used to constrain the properties of
mini-charged particles.  We outline how this test is constructed, and
present new constraints on mini-charged particles from observations of
the Sunyaev-Zel'dovich effect in the Coma cluster.
    \vspace{3mm}

    \end{abstract}
    \maketitle

\maketitle

\section{Introduction}
Precision measurements of the Cosmic Microwave Background (CMB)
radiation have, in recent years, enormously advanced our understanding
of the origins, content and structure of our universe.  Measurements
of secondary anisotropies induced, not at the surface of last
scattering ($z\approx 1100$), but in the more recent universe ($z\sim
\mathcal{O}(1)$) also provide detailed information about the late time
evolution of the universe enabling, for example,  measurements of the Hubble constant
through the Sunyaev-Zel'dovich (SZ) effect (for a review see \cite{2002ARA&A..40..643C}), and measurements of the
properties of dark energy through the late time Integrated Sachs-Wolfe
effect (ISW) \cite{Sachs:1967er,Giannantonio:2008zi}.

These precision measurements can also be used to test `new physics',
including the existence of new light, weakly interacting particles
if they influence the propagation of photons. In this article we
focus in particular on mini-charged particles (MCPs).  MCPs  are
particles with a small and not necessarily quantized charge. Such
particles arise naturally in extensions of the Standard Model which
contain at least one additional U(1) hidden sector gauge
group~\cite{Holdom:1985ag,Bruemmer:2009ky}. The gauge boson of this
additional U(1) is known as a hidden photon, and hidden sector
particles, charged under the hidden U(1) get an induced electric
charge proportional to the small mixing angle between the kinetic
terms of the two photons. In string theory such hidden U(1)s and the
required kinetic mixing are a generic
feature~\cite{Abel:2006qt,Abel:2008ai,Dienes:1996zr,Lukas:1999nh,Lust:2003ky,Abel:2003ue,Blumenhagen:2005ga,Blumenhagen:2006ux,Mark}.
Hidden photons are not {\it necessary} however to explain
mini-charged particles, and explicit brane world scenarios have been
constructed \cite{Batell:2005wa} where MCPs arise without the need
for hidden photons.

The existence of MCPs would lead to the decay of photons in
the presence of electric or magnetic fields \cite{Gies:2006ca,Ahlers:2006iz}.  This has lead to
searches for MCPs in high precision optical experiments (BFRT \cite{PhysRevD.47.3707},
PVLAS \cite{zavattini:110406}, Q\&A \cite{Chen:2006cd}, BMV \cite{Rizzo} and OSQAR~\cite{Pugnat:2005nk}) where
  a laser beam is passed
through a transverse magnetic field and the real and virtual production of MCPs leads to rotation and ellipticity
of the polarization of the beam.  This signal
differs depending on whether or not the model under examination includes
hidden photons.  In addition  the presence of hidden photons can  lead
to more exotic effects, such as light-shining-through-walls \cite{Ahlers:2007rd,Ahlers:2007qf}.

For a wide range of parameters, however,  more stringent constraints
on MCPs come from observations in astrophysics and cosmology
\cite{Davidson:1993sj,Davidson:2000hf}.  In particular  extensions of the Standard
Model which include MCPs must be in agreement with the bounds of Big
Bang Nucleosynthesis (BBN), and must not lead to overly fast cooling
of white dwarf and red giant stars.  However it has been shown
\cite{Masso:2006gc} that in models containing more than one hidden
photon, where at least one of the  hidden photons is massless, the
bounds obtained in settings with high density/temperature can be
considerably relaxed. Most prominently this affects bounds from
energy loss considerations in stars. Therefore alternative
constraints obtained in low density/temperature environments are of
particular interest~\cite{Melchiorri:2007sq,Ahlers:2009kh}. For this
we turn to observations of the CMB; the light we observe  from the
CMB has traveled solely through low density/temperature
environments, and therefore constraints on MCPs derivable from the
CMB also apply to those models  which avoid the stellar evolution
bounds. These constraints would  be of direct relevance for upcoming
laboratory searches for MCPs which are typically performed under
vacuum conditions. The light from the CMB passes through  magnetic
fields in the neighborhood of galaxy clusters, where real and
virtual MCPs are produced exactly as in laboratory experiments.  The
anisotropies induced by such interactions  contribute to the
standard late time CMB anisotropies.  In this article we show that
observations of these effects can be used to constrain new regions
of the MCP parameter space.

Cluster magnetic fields are well understood on distance scales at
which the  SZ effect dominates over the ISW effect, but little is
known about magnetic fields in the
largest structures in the universe. Therefore in this article we
mainly focus on an MCP contribution to the SZ effect.  When photons from the CMB pass through  galaxy clusters there is a
small probability that they will interact with an energetic electron
in the plasma of the intracluster medium.  If this happens the photons
can be Thomson scattered up to higher energies, distorting the CMB
spectrum.  This is the Sunyaev-Zel'dovich (SZ) effect
\cite{Zeldovich:1969ff,Sunyaev:1972eq}.  The temperature distortions
induced in the CMB have  the form
\begin{equation}
\frac{\Delta T}{T} =f\left(\frac{\omega}{T_{CMB}}\right)\int\frac{n_e
T_e \sigma_T}{m_e}\;dl,
\end{equation}
where $\omega=2\pi \nu$ is the photon energy,  $T_{CMB}$ the CMB temperature, $n_e$ the
electron number density in the plasma, $T_e$ the electron temperature,
$\sigma_T$ the Thomson scattering cross section and $m_e$ the mass of
the electron\footnote{We are using units $k_B=\hbar=c=1$.}.
The function $f(x)$ describes the frequency dependence
of the SZ effect and in the non-relativistic and Rayleigh-Jeans
($\omega\ll T$) limits $f(x)\rightarrow -2$.  The integral is along a
line of sight through a cluster.
A typical galaxy cluster is expected to induce temperature anisotropies of the
order $10^{-4}$ in the CMB spectrum.  Photons are lost in the long
wavelength part of the CMB spectrum $\nu \lesssim 218 \mbox{ GHz}$ and
there is an increase in the power of the spectrum at higher
frequencies.  There are now a large number of high quality
measurements of the SZ effect for a variety of clusters. The physics of the SZ
effect is reviewed in \cite{Birkinshaw199997} and the current
observations are reviewed in \cite{2000cucg.confE..43C}.

 Constraints on MCPs from
observations of the
CMB have also been derived from processes where two CMB photons
annihilate into two MCPs \cite{Melchiorri:2007sq} and in the region of
parameter space where MCPs do not decouple from the acoustic
oscillations of the baryon-photon plasma at recombination \cite{Dubovsky:2003yn}.  Other cosmological
probes of MCPs have also been considered, including their effect on
the propagation of light from type Ia supernovae
\cite{Ahlers:2009kh}.  Modifications of the SZ effect by chameleonic
axion-like-particles have also been discussed \cite{Davis:2009vk}.

The outline of this article is as follows.  In Section \ref{sec:optics} we
describe the effect of MCPs on the propagation of photons through
magnetic fields.  We solve the equations of motion and compute the
survival probability for photons.   In Section \ref{sec:SZ} we show
how measurements of the SZ effect can be used to constrain MCPs, in
Section \ref{sec:coma} we give the constraints on MCP models that come from
 observations of the Coma cluster and in Section \ref{sec:hyper} we
 discuss the relevance of these constraints to hyperweak gauge
 interactions in the LARGE volume scenario of string theory.  Section \ref{sec:ISW} describes how,
in the future, observations of the ISW effect may also be used to
constrain MCPs, and we conclude in Section \ref{sec:conc}.

\section{Optics with MCPs and hidden photons}
\label{sec:optics}
Photon propagation in the presence of mini-charged particles can be  studied
in the Holdom model \cite{Holdom:1985ag}  starting with the most general
Lagrangian
\be
{\cal L} = -\frac{1}{4}\tilde F_{\mu\nu}\tilde F^{\mu\nu}-\frac{1}{4}\tilde B_{\mu\nu}\tilde B^{\mu\nu}
-\frac{\sin\chi}{2}\tilde B_{\mu\nu}\tilde F^{\mu\nu} +
\tilde e j_{\rm em}^\mu \tilde A_\mu + e_h j_{\rm h}^\mu \tilde B_\mu,
\ee
describing  visible sector photons $\tilde A_\mu$, hidden sector
photons $\tilde B_\mu$, and visible and hidden sector matter fields,
written here as the currents $j_{\rm em}^\mu$ and $j_{\rm h}^\mu$
respectively. The visible and hidden  photons have field strength
tensors $\tilde F_{\mu\nu}$ and $\tilde B_{\mu\nu}$ respectively, and $\chi$ controls the
strength of the kinetic mixing between the photon fields.
The visible and hidden sector gauge couplings are $\tilde e$ and $e_h$.

The following change of variables
\be
\label{shift}
\tilde A = \frac{1}{\cos \chi}A \quad ; \quad \tilde B = B- \tan \chi A,
\ee
diagonalizes the kinetic part of the Lagrangian, which can then be written as
\be
\label{lag}
{\cal L} = -\frac{1}{4} F_{\mu\nu}F^{\mu\nu}-\frac{1}{4} B_{\mu\nu} B^{\mu\nu} +
e j_{\rm em}^\mu A_\mu + e_h j_{\rm h}^\mu B_\mu + \epsilon e j^{\mu}_{\rm h} A_\mu,
\ee
with $\epsilon = (e_h/e)\tan \chi \quad$ and $\quad
e = \tilde e/\cos \chi$.
The last term of (\ref{lag}) gives an effective charge under the visible sector gauge group to the hidden
matter.
If either $\chi$ or $e_{\rm h}$ are small the effective charge of the
hidden matter has a naturally small value $|\epsilon|\approx e_{h}\chi/e\ll 1$.

\begin{figure}[tbp]
\centering
{
\psfragscanon
\psfrag{a}[][l]{$\lambda=\frac{3}{2}\frac{\omega}{m_\epsilon}\frac{\epsilon e B}{m_\epsilon^2}$}
\psfrag{b}[][l][0.8]{}
\psfrag{c}[][l][0.8]{}
\psfrag{d}[][l][0.8]{\vspace{1cm}\hspace{1cm}  $\begin{array}{c}\ |{\rm Re} (\mu^2)| $ - - - $    \\ |{\rm Im} (\mu^2)|\  $-----$  \end{array}$}
\includegraphics[width=8cm]{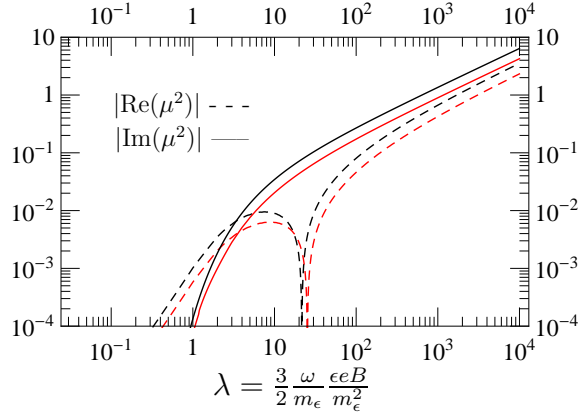}
}
\caption{\small The absolute value of the real (solid) and imaginary (dashed) parts of the MCP-induced mass $\mu^2$
are shown for photon polarization parallel (black) and perpendicular (red) to the magnetic field direction.
The MCP particle is a Dirac spinor with mass $m_\epsilon$ and electric charge $\epsilon$. The scalar case is
very similar. They only depend on the adiabatic parameter $\lambda$.
The imaginary part of $\mu^2$ is always negative while the real part is negative
for $\lambda \lesssim 20$ and becomes positive for larger
values.}
\label{fig:mu2}
\end{figure}

In the presence of a strong transverse magnetic  field the hidden matter contributes to the refraction
and absorption coefficients of photons and hidden photons
\cite{Gies:2006ca} through the complex refractive index,
$\epsilon^2e^2\Delta N_i(\epsilon e
\mathbf{B},m_{\epsilon})=\epsilon^2e^2(\Delta
n_i+i\frac{1}{2\omega}\kappa_i)=n_i-1$, for a photon of frequency
$\omega$ and an MCP with mass $m_{\epsilon}$. $i=\perp,\parallel$
labels the photons polarized parallel and perpendicular to the
direction of the magnetic field. The real parts, $\Delta n_i$, are the
refractive indices and the imaginary parts, $\kappa_i$,  are
the absorption coefficients due to the production of MCPs.  Full expressions for $\Delta N_i$
are given in Refs.~\cite{Erber:1966vv,Tsai:1974fa,Gies:2006ca,Ahlers:2006iz}.
The  equations of motion derived from the Lagrangian (\ref{lag}) are
\begin{equation}
\label{eom}
\left[(\omega^2+\partial_z^2)\left(\begin{array}{cc}
1 & 0\\
0 & 1
\end{array}\right)-\left(\begin{array}{cc}
\OP^2 + \mu^2\chi^2 & - \mu^2\chi\\
- \mu^2\chi & \mu^2
\end{array}\right)\right]\left(\begin{array}{c}
A\\
B
\end{array}\right)=0,
\end{equation}
where we have defined
\be
\mu^2 = -2\omega^2e_h^2\Delta N_i
\ee
and $\tan(\chi)\to \chi$. $\OP^2=4\pi^2\alpha n_e/m_e$ is the plasma
frequency depending on the fine structure constant, the mass of an
electron, $m_e$, and the number density, $n_e$, of free electrons in a plasma.
The effective mass $\mu$,  normalized by the MCP mass, depends only on
 the polarization of the light  and on the
adiabatic parameter
\be
\lambda = \frac{3}{2}\frac{\omega}{m_\epsilon}\frac{\epsilon e B}{m_\epsilon^2}.
\ee
The dependence  of $\mu^2/m^2_\epsilon$ on $\lambda$ is shown in
Figure ~\ref{fig:mu2}.

Solving the equations of motion, the propagating eigenstates are
\be
\label{prop}
V_+(t,z)    = \left(\begin{array}{c}   1  \\   - a   \end{array} \right) e^{i (\omega t - k_+ z)}   \quad ; \quad
V_-(t,z)    =  \left(\begin{array}{c}   a  \\   1   \end{array} \right) e^{i (\omega t - k_- z)},
\ee
where the momenta are
\bea
k_\pm &=& \sqrt{\omega^2-m_\pm^2    }\simeq \omega -\frac{m_\pm^2}{2\omega}\equiv \omega - \Delta_\pm,          \\
\label{mpm}
2 m_\pm^2   &=& \OP^2+\mu^2(1+\chi^2)\pm \sqrt{(\OP^2-\mu^2(1-\chi^2))^2+4\mu^4\chi^2},
\eea
and
\bea
\label{a}
a=\frac{\mu^2}{m_+^2-\mu^2}\chi.
\eea
Therefore a state which is  purely photon-like initially at $z=0$ evolves as
\be
\label{Vz}
V(t,z) = \left(\frac{1}{1+a^2}V_+(0,0)e^{i\Delta_+ z} + \frac{a}{1+a^2}V_-(0,0)e^{i\Delta_- z}\right)e^{i\omega(t-z)}.
\ee
The photon survival amplitude is the first component of this vector
and, from this,  the photon survival probability is
\be
\label{genprob}
P_{\gamma\to\gamma}(z)=\left|\frac{1}{1+a^2}\right|^2\Bigl(e^{-2 {\rm Im}\Delta_+z} + |a|^4e^{-2 {\rm Im}\Delta_-z} +2{\rm Re} \{a^2 e^{i (\Delta_--\Delta^*_+)z}\}\Bigr)
\ee
The last term inside  the bracket in (\ref{genprob}) is oscillatory and can be neglected when
the phase is large  $\phi = {\rm Re}\{\Delta_--\Delta^*_+\} z \gg 1$,
corresponding to situations where  a large number of oscillations
occur within the propagation length $z$.

The expression for the survival probability (\ref{genprob}) breaks down
when $a^2=-1$ or, equivalently, when
\be
\label{resonance*}
\mu^2= \frac{\OP^2}{(1\pm i \chi)^2}.
\ee
This is the point in phase space where the  photon and hidden photon
are exactly degenerate.
As the imaginary and real parts of $\mu^2$ satisfy
\be
\label{caca}
{{\rm Im}}\{ \mu^2\} > {{\rm Re}}\{ \mu^2\} \hspace{2cm} {\rm for}  \hspace{1.5cm} {{\rm Re}}\{ \mu^2\}>0 \ ,
\ee
the condition (\ref{resonance*}) can only be satisfied for values of $\chi\sim {\cal O}(1)$.
In this paper we  restrict ourselves to considering small values of
$\chi$, which are not only  more realistic from the theoretical point of view but also are not excluded by laboratory experiments.
Therefore we are always far from the resonance.

In the small mixing regime   $\chi \ll 1$ a simpler formula for the photon survival probability  (\ref{genprob})
can be obtained.
Expanding (\ref{mpm}) and (\ref{a}) around $\chi=0$ we find
\bea
m_+^2   &=& \OP^2\(1+a\chi\) \quad ; \quad  m_-^2 = \mu^2\(1-a\chi\)            \\
a       &=& \frac{\mu^2}{\OP^2-\mu^2}\chi
\eea
So that the photon survival probability becomes
\begin{equation}
P_{\gamma\to\gamma}(z)=1-2{{\rm Re}}\{a^2\}-\chi\frac{z\omega_P^2}{\omega}{{\rm Im}}\{a\}+2{{\rm Re}}\{a^2e^{-iz(\omega_P^2-\mu^2)/2\omega}\}+\mathcal{O}(\chi^4).
\label{smallchi}
\end{equation}
The last term in this expression is exponentially damped and when the
distances under consideration satisfy $z{{\rm Im}}\{\mu^2\}\gg2\omega$ it
can be safely neglected.

As mentioned in the introduction the inclusion of the  hidden photon
is not necessary for the existence of MCPs but it  certainly
provides one of the few natural theoretical explanations of the small mini-charges.
Use of the framework described above  imposes no restriction
on the origin of MCPs because the hidden photon field can be consistently decoupled by
formally sending $e_h\to 0$ whilst keeping $\epsilon$ constant. This can be seen
directly from the Lagrangian (\ref{lag})  where  the only term that
connects the hidden photon to the other fields
 is proportional to $e_h$\footnote{This decoupling can also be seen in the matrix form of
the equations of motion (\ref{eom}).
As $\mu^2$ equals $e_h^2$ multiplied by some function $f(\epsilon)$ the $A-A$ matrix element
is $\propto \chi^2 e_h^2 f(\epsilon)=\epsilon^2e^2 f(\epsilon)$ and will stay constant in the decoupling limit,
however the $A-B$ element is $\propto \chi e_h^2 f(\epsilon)=e_h \epsilon e f(\epsilon)$ and vanishes as
$e_h\to 0$.}.
Note that $m_+^2$ is a constant in this limit but $\chi \mu^2$ and
$\mu^2$ vanish and therefore  the mixing parameter $a$ also
tends to zero.
When the hidden photon decouples the photon survival probability
becomes simply
\begin{equation}
P=e^{-2 {\rm Im}\Delta_+z}.
\end{equation}

\section{Using the Sunyaev-Zel'dovich effect to constrain MCPs}
\label{sec:SZ}
As discussed in the introduction, magnetic fields exist in clusters of
galaxies.  As light from the CMB traverses these fields the properties
of this light can be affected by the real and virtual production of
MCPs.  Clusters of galaxies are some of the largest objects in the
  universe; a typical cluster contains $\sim 10^3$ galaxies in a
  region $\sim 2$ Mpc in radius. The magnetic fields of these
  clusters are relatively well understood \cite{Carilli}, and it is
  common to model these magnetic fields as being made up of a large
  number of equally sized magnetic domains.  Within each domain the
  magnetic field is constant, and the magnitude of the magnetic field
  strength is constant over the whole cluster, but within each domain
  the direction of the magnetic field vector is essentially a random
  variable.  Photons passing
through such clusters may interact with the cluster magnetic
field and  convert into real or virtual MCPs.  This loss of photons would
look like a contribution to the SZ effect of the form
\begin{equation}
\frac{\Delta T}{T}=\frac{1-e^{-x}}{x}\frac{\Delta I}{I_0},
\label{MCPSZ}
\end{equation}
where $I_0$ is the photon flux
from the CMB, $\Delta I$ is the flux decrement due to MCPs and
$x=\omega/T_{CMB}$, and $ T_{CMB}$ is the
temperature of the CMB radiation today.
The best constraints would come from a cluster for which both the SZ effect
and the properties of its magnetic field have been directly measured.
This is uniquely
the case for the Coma cluster (Abell 1656) which lies at a redshift
$z=0.023$.   The properties of the Coma cluster will be discussed
further in Section \ref{sec:coma}.

In order to constrain the MCP contribution to the SZ effect we must
compute the flux deficit induced as  the photons propagate through a large
number of randomly oriented magnetic domains.    As the CMB is a {\it very} non
compact source,  light from the CMB  takes many
different paths through the random magnetic field of a
galaxy cluster and we need only compute  the
effects of MCP production averaged over this large class of paths.

We will assume that the size of a magnetic domain is sufficiently
large that the final term in the photon survival probability
(\ref{smallchi}) can be neglected.  This is consistent whenever
${{\rm Im}}\{\mu^2\}\gg2\omega/L$, where $L$ is the size of the
domain.    Then the photon
survival probability can be written as $P_i(z)
=1-p_i-q_iz$ for $i=\parallel,\perp$, where $p=2{{\rm Re}}\{a^2\}$,
$q=\chi\omega^2_P{{\rm Im}}\{a\}/\omega$.  $a$ is given by (\ref{a}). To evolve the system through the cluster we will need to
average over the two angles $\theta_n$ and $\psi_n$ that determine the
direction of the magnetic field in the n-th domain. The average over
$\psi_n$, the angle of inclination of the magnetic field to the
direction of motion of the photons, we will absorb into a
conservative order of magnitude estimate for $B$.  Letting $\theta_n$ be the
orientation of the magnetic field in the plane perpendicular to the
direction of motion of the photons, the
photon components at the  start  of the $(n+1)$-th domain are related to those at the start of the $n$-th
domain by\footnote{In principle the propagation is described by a $4\times4$ matrix. However, after the initial damping only the $V_{+}$ eigenmodes
survive and we can use effectively a $2\times 2$ description.}
\begin{eqnarray}
\left(\begin{array}{c}
A_x\\
A_y
\end{array}\right)_{n+1}&=&\left[ \left(1-\frac{\delta_{n1}\langle p\rangle+\langle
    q\rangle L}{2}\right)\mathbb{1}\right.\nonumber\\
& &+\left.\frac{\delta_{n1}dp +dqL}{2}\left(\begin{array}{cc}
\cos 2\theta_n & \sin2\theta_n\\
\sin2\theta_n  & -\cos2\theta_n
\end{array}\right)\right]\left(\begin{array}{c}
A_x\\
A_y
\end{array}\right)_{n}
\end{eqnarray}
where $\langle q\rangle =(q_{\bot}+q_{\parallel})/2$ and
$dq=(q_{\bot}-q_{\parallel})/2$, with equivalent definitions for
$\langle p\rangle$ and $dp$.
Assuming that, on average, $\cos^2\theta_n=\sin^2\theta_n=1/2$ and
$\cos\theta_n=\sin\theta_n=0$ it can be shown that after
passing through $N$ magnetic domains the average photon flux  $I_{N}$,  is
related to the initial photon flux, $I_0$, by
\begin{equation}
I_{N}=I_0\Bigl(1-\langle p\rangle-N\langle q\rangle L +\mathcal{O}(N^2\langle q \rangle^2)\Bigr).
\label{flux}
\end{equation}
Combining equations (\ref{MCPSZ}) and (\ref{flux}) we can constrain the CMB
temperature anisotropies induced in MCP models to be less than those
of the SZ effect.

In the general case we compute the photon survival probabilities
 numerically, but there are two limits which can be
understood analytically.
If the plasma frequency is sufficiently large $\OP^2\gg |\mu^2|$, the
imaginary part of the mass of the  photon like state (+)  is well
approximated  by ${{\rm Im}}\{m_+^2\} =\chi^2 {{\rm Im}}\{ \mu^2\} = -\epsilon^2 e^2 \kappa \omega$.
This regime is equivalent to the  decoupling of the hidden photon
discussed in Section \ref{sec:optics}
($e_h\to 0$, $\mu^2\to 0$ but $\chi^2\mu^2=\mbox{constant}$).
In this regime the mixing angle $a$ is suppressed not only by $\chi$
but also by the ratio $\mu^2/\OP^2$ and  the photon survival probability is simply
\be
P(\gamma\to \gamma) =  1-z\epsilon^2e^2\kappa + {\cal O}(a^2)  .
\label{nohp}
\ee
So $\langle p\rangle=0$ and $\langle q\rangle=\epsilon^2e^2\langle\kappa\rangle$ in (\ref{flux}).

In the opposite limit, $|\mu^2|\gg \OP^2$, the situation also simplifies, recovering the expressions derived
in \cite{Ahlers:2007rd}. The mixing angle is only suppressed by $\chi$ and is
real in the limit $\OP\to 0$. The mass of the photon like state (+)
still has an imaginary part, but this is suppressed by the small quantity $\OP^2/|\mu^2|$,
\be
 {{\rm Im}}\, \{m^2_+\}  \simeq  \chi^2{{\rm Im}}\, \{\mu^2\}  \frac{\OP^4}{|\mu^2|^2} = \omega\epsilon^2 e^2\kappa \frac{\OP^4}{|\mu^2|^2}.
\ee
In this limit the photon survival probability is
\be
P(\gamma\to \gamma) \simeq 1-2\chi^2-4\chi^2\frac{\omega^2_P{{\rm Re}}\{\mu^2\}}{|\mu^2|^2}-\chi^2\frac{z\omega^2_P{{\rm Im}}\{\mu^2\}}{\omega|\mu^2|^2}+\mathcal{O}\left(\frac{\omega_P^4}{|\mu^2|^2}\right).
\label{Pnoomega}
\ee
In the limit $\OP\to 0$ the photon survival probability becomes constant
\be
\lim_{\OP\to 0}P(\gamma\to \gamma) \simeq 1-2\chi^2 .
\label{justhp}
\ee
So $\langle p\rangle =2\chi^2$ and  $\langle q\rangle =0$ in (\ref{flux}).
This can be understood by considering equation
(\ref{Vz}). In the $\OP\to 0$ limit, $a\to -\chi$ and an initial photon state
is mainly $V_+$ with a very small (proportional to  $\chi^2$) component of $V_-$.
In this case, the $V_-$ state is the original hidden photon $\tilde{B}_\mu$  which,
by definition, is the state that couples to the hidden sector
particles.
Therefore only the $V_-$ component of the initial state can be damped by pair production of
MCPs and after traveling  distances $z\gtrsim 2\omega\ln (\chi)/{{\rm Im}}\{\mu^2\}$  the final state
is  $V_+/(1+\chi^2)$.  Squaring the amplitude of this gives the photon
survival probability (\ref{justhp}).

\subsection{Constraints from the Coma cluster}
\label{sec:coma}
 The most detailed information
about the strength and structure of the magnetic fields in clusters of
galaxies typically comes from measurements of Faraday rotation of
light at radio frequencies.  The presence of a magnetic field in an
ionized plasma, such as the intracluster medium, sets a preferred direction for the gyration of
electrons, leading to a difference in the index of refraction for left
and right circularly polarized radiation as it passes through the
plasma.  This is equivalent to a rotation of
the plane of polarization of linearly polarized light, known as the
Faraday rotation, which depends on the thermal electron density and
the magnetic field strength.  By taking Faraday rotation measurements
of an entire cluster, not only the magnetic field strength, but also
its coherence length, can be estimated.  It has been shown
\cite{Gies:2006ca,Ahlers:2006iz}, however, that the
interactions of photons and mini-charged particles in a transverse magnetic field also lead to the rotation of polarization, and
these  effects are expected to be strongest at low frequencies.
Therefore  it is unclear whether the magnetic field strengths calculated from Faraday rotation
measurements can be reliably used in the computation of CMB
temperature anisotropies due to MCPs.

Faraday rotation is not, however, the only way to estimate the
magnetic field strength of a cluster.  For the Coma cluster a hard
X-ray flux has been measured exceeding the thermal emission.  This has
its origin in the inverse Compton scattering of photons  by relativistic
electrons which are accelerated by the magnetic field of the galaxy cluster. This inference of the magnetic field is not based on subtle
polarization measurements and therefore we expect it to be essentially free of
contamination by MCPs.  Hard X-ray
observations of the Coma cluster imply a magnetic field
strength of $0.16\times 10^{-10}\mbox{ T}$ \cite{1999A&A...344..409E,1999ApJ...520..529S}, roughly an order of magnitude
smaller than that inferred from Faraday rotation measurements \cite{1990ApJ...355...29K}.
Whilst, within the Standard Model, it is thought that these two sets of observations can be
reconciled by more realistic cluster models  \cite{Brunetti}.  The
size of a magnetic domain can be estimated from images of the Faraday
rotation of the Coma cluster, $L\approx 10 \mbox{ kpc}$,
\cite{1990ApJ...355...29K} and the size of these correlations
will not be affected by mixing with MCPs.

We note that it  would be possible to derive constraints on MCP models by requiring that
the
MCP induced rotation of linearly polarized light at radio frequencies
be less than the measured Faraday rotation \cite{1990ApJ...355...29K}.  However these
constraints are subsumed by the constraints coming from measurements  of
the SZ effect, which we focus on for the remainder of this article.

To compute the contribution of MCPs to the SZ effect we also need to
know the electron density in the intracluster plasma.  This can be inferred, for the Coma cluster, from soft X-ray
measurements taken during  the ROSAT all sky survey
\cite{1992A&A...259L..31B}.  From these observations it is inferred that  $n_e=2.89\times 10^{-3}\mbox{ cm}^{-3}$
in the core of the cluster and  $n_e\approx 1 \times
10^{-3}\mbox{ cm}^{-3}$ away from the core.  This corresponds to a
plasma frequency of $\omega_P\approx 10^{-12} \mbox{ eV}$.

The SZ effect of the Coma cluster has been measured precisely in
a number of frequency bands \cite{1538-4357-598-2-L75}, as shown in
Table \ref{table:obs}.
\begin{table}[ht]

\centering 
\begin{tabular}{|c| c| c| c|} 
\hline
Instrument &  & Temperature Decrement &Observational  Frequency\\[0.5ex] 
\hline 
OVRO & \cite{1538-4357-449-1-L5} &$\Delta T = -520\pm 83 \mbox{
  $\mu$K}$ &32.0 GHz \\
WMAP & \cite{0067-0049-148-1-97} & $\Delta T = -240\pm 180 \mbox{
  $\mu$K}$ & 60.8 GHz\\
WMAP & \cite{0067-0049-148-1-97} & $\Delta T = -340\pm 180
\mbox{ $\mu$K}$ & 93.5 GHz\\
MITO & \cite{1538-4357-574-2-L119} &$\Delta T = -184\pm 39
\mbox{ $\mu$K}$ & 143 GHz\\
MITO & \cite{1538-4357-574-2-L119} & $\Delta T = -32\pm 79
\mbox{ $\mu$K}$ & 214 GHz\\
MITO & \cite{1538-4357-574-2-L119} & $\Delta T = 172\pm 36
\mbox{ $\mu$K}$ & 272 GHz\\ [1ex]
\hline 
\end{tabular}
\caption{SZ observations of the Coma Cluster}
\label{table:obs} 
\end{table}	
For each observation we constrain the temperature decrement due to
production of MCPs with the largest measured temperature decrement
within $2\sigma$ error bars.  All the observations detailed above
give constraints on the parameters of the MCP model of the same order
of magnitude, with those  by MITO at 214 GHz, being the most
constraining.  The numerical bounds quoted below come from this measurement.

To calculate the constraints on MCPs from the SZ measurements of the
Coma cluster we  assume that in most of the volume of the
Coma cluster  $B\approx 1\times  10^{-11}\mbox{ T}$, which estimate
includes an averaging over the angles $\psi_n$,  and that the size
of a magnetic
domain is $L\approx 10\mbox{ kpc}$.  The cluster magnetic fields
extend out to a radius of at least  1 Mpc, so we  assume that light from
the CMB traverses approximately 100 domains.  We take the average plasma
frequency to be $\omega_P=10^{-12}\mbox{ eV}$.  The MCP induced
temperature anisotropies (\ref{MCPSZ}) at $\nu \sim 214\mbox{ GHz}$ are,
from (\ref{flux}), constrained to be
\begin{equation}
\frac{\Delta T}{T}=0.25\Bigl(\langle p\rangle+NL\langle q\rangle\Bigr) < 7.0 \times 10^{-5}.
\end{equation}
Our assumption that the last term in (\ref{smallchi}) can be safely
neglected when calculating the survival probability in each domain of
the Coma cluster means that our constraints are valid for
${{\rm Im}}\{\mu^2\}\gg2\omega/L$, which for observations at 214 GHz implies
${{\rm Im}}\{\mu^2\}\gg1.1\times 10^{-30}\mbox{ eV}^2$.

\begin{figure}[tbp]
{ 
\psfragscanon
\psfrag{a}[][l]{$m_\epsilon$ [eV]}
\psfrag{b}[][l]{$\epsilon$}
\psfrag{ss1}[][l][0.7]{$e_h=10^{-5}$}
\psfrag{ss2}[][l][0.7]{$e_h=10^{-4}$}
\psfrag{ss3}[][l][0.7]{$e_h=10^{-3}$}
\psfrag{ss4}[][l][0.7]{$e_h=3\times 10^{-3}$}
\psfrag{ss5}[][l][0.7]{$e_h=10^{-2}$}
\psfrag{ss6}[][l][0.7]{$e_h=0.1$}
    \includegraphics[height=11cm]{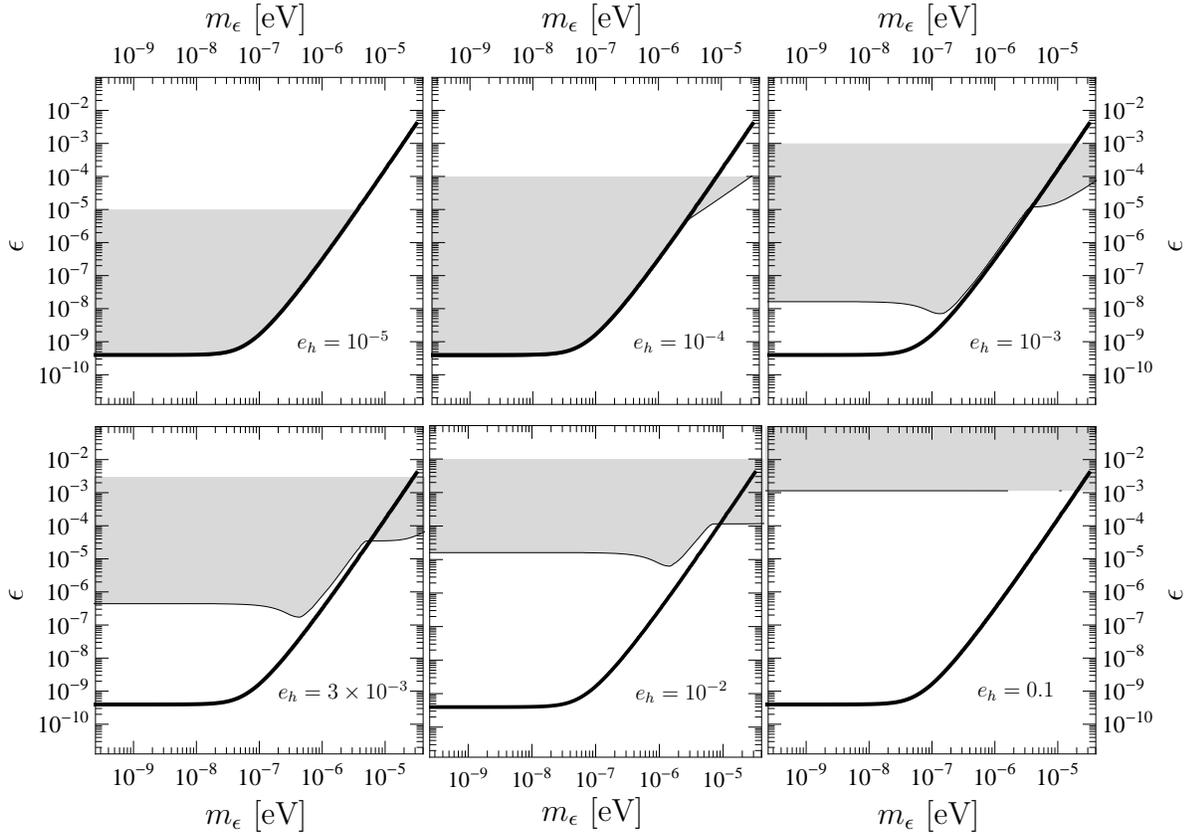}   }
\caption{The constraints of the SZ effect on MCPs.  The solid black
  line shows the upper bound in the mini-charge, MCP mass phase space on models which contain MCPs but no
  hidden photons (we consider a Dirac fermion, but scalars are very similar).  The gray region shows the excluded region for
  models which include hidden photons.  The different plots show
  different values of the hidden sector gauge coupling, given in units
  of the electric charge.}
\label{fig:caca}
    \end{figure}

The constraints this imposes on the charge and mass of  MCPs  are shown
in Figure \ref{fig:caca}.  The upper bound  on models without hidden
photons is shown as a solid line.  The gray regions are those
excluded  for models which include both hidden sector photons
and MCPs. The upper bound on these regions  comes from requiring
that $\chi\equiv\epsilon e /e_h < 1$.  The excluded region in the $(m_{\epsilon},\epsilon)$
plane  for  models with hidden photons varies significantly
with $e_h$.  For small values of the hidden sector gauge coupling the
constraints  are indistinguishable from those of the pure MCP case. This is the $|\mu^2|\ll\OP^2$
limit discussed in the previous section.  For larger gauge couplings the
constraints on the mini-charge are much weaker because in this region
of the parameter space only the hidden photon component of the initial
state is  damped by the production of MCPs. This corresponds to
the  $|\mu|^2\gg \omega_P^2$ limit.

In certain regions of the parameter space it is possible to understand
these limits analytically.  When the adiabatic parameter is large
$\lambda\gg1$ an analytic expression for the complex refractive index exists.
  For light at 214 GHz passing through the magnetic fields of the  Coma
cluster large adiabatic parameter corresponds to
\begin{equation}
\epsilon^{1/3} \gg 1.1\times 10^4
\left(\frac{m_{\epsilon}}{\mbox{eV}}\right).
\end{equation}
and this  corresponds to the region on the left of
Figure \ref{fig:caca}  where the constraints are independent of the
MCP mass.

When the adiabatic parameter is large and $|\mu|^2\ll
\omega_P^2$,  corresponding to $ e_h\epsilon^{1/3}\ll 7\times
10^{-8}$,  the hidden photons effectively decouple.  The
photon survival probability is given by (\ref{nohp}) and measurements of the SZ effect in the
Coma cluster constrain
\begin{equation}
\epsilon < 4\times 10^{-10}.
\end{equation}
In the alternative limit   $|\mu|^2\gg \omega_P^2$ the photon survival
probability is  given by (\ref{justhp}) and measurements of the SZ effect constrain
\begin{equation}
\chi<1\times 10^{-2},
\end{equation}
or equivalently
\begin{equation}
\epsilon<3\times 10^{-2}e_h.
\end{equation}
\subsection{Hyperweak hidden gauge couplings}\label{sec:hyper}
Above we have seen that our bounds are strongest for mini-charged particles without hidden photons and for very small hidden sector gauge couplings.
Although the first case is interesting in itself, let us also briefly motivate the case of small hidden sector gauge couplings.
Indeed we will see below that our bounds start to penetrate a theoretically very interesting region.

Hyperweak gauge interactions~\cite{Burgess:2008ri} are a typical feature in so-called LARGE volume scenarios in string theory.
In LARGE volume scenarios the string scale $M_{s}$ is related to the Planck scale $M_{P}$ (and the string coupling $g_{s}$) via
\begin{equation}
M^2_{P}=\frac{4\pi}{g^2_{s}}{\mathcal{V}}M^{2}_{s},
\end{equation}
with a LARGE volume ${\mathcal{V}}=V_{6}M^{6}_{s}$ of the six compactified dimensions.
Due to the LARGE volume the string scale can now be much lower than the Planck scale.
For example for a volume of the order of
${\mathcal{V}}\sim 10^{12}$ one obtains a string scale $M_{s}\sim 10^{11}\,{\rm GeV}$. This scale is particularly interesting in scenarios where
a supersymmetry breaking of size $\sim M^{2}_{s}$ is mediated to the Standard Model by gravity resulting in masses for the superpartners of the order
of $\sim M^{2}_{s}/M_{P}\sim 1\,{\rm TeV}$.
For an even larger volume ${\mathcal{V}}\sim 10^{27}$ the string scale itself could be as low as a TeV.

In the LARGE volume scenarios gauge groups live on D7 branes with $7+1$ dimensions. The extra 4 space dimensions are removed by wrapping the brane
around a cycle in the extra dimensions. The gauge coupling is then given by
\begin{equation}
g^{2}=\frac{2\pi g_{s}}{{\mathcal{V}}_{4}}\sim \frac{2\pi g_{s}}{{\mathcal{V}}^{\frac{2}{3}}},
\end{equation}
where ${\mathcal{V}}_{4}=V_{4}M^{4}_{s}$ is the volume of the compactified 4 extra dimensions of the brane.
In the last step we have assumed that the brane extents through
most of the LARGE volume and the latter has roughly the same extent in all directions.
In this case the gauge coupling will be very small. (The Standard Model gauge groups live on smaller cycles or singularities and accordingly have
larger gauge couplings $\sim 1$.)
Typical values for these hyperweak gauge couplings are of the order
\begin{equation}
e_{h}\sim \bigg\{\begin{array}{clcc}
          10^{-4}& \sim  10^{-3}\,e & {\rm for} & {\mathcal{V}}\sim 10^{12} \\
          10^{-10} &\sim  10^{-9}\, e& {\rm for} & {\mathcal{V}}\sim 10^{27}
        \end{array}.
\end{equation}
This is exactly the regime where our bound is most constraining.

Finally, let us also take a brief look at the kinetic mixing in these scenarios (cf. \cite{Mark} for details).
An estimate of the kinetic mixing between a visible sector (non-hyperweak) U(1) and a hidden hyperweak U(1) gauge group
yields,
\begin{equation}
\chi\sim \frac{e e_{h}}{6\pi^2}.
\end{equation}
Accordingly we find for the mini-charge
\begin{equation}
|\epsilon|=\left|\frac{e_{h}\chi}{e}\right|\sim \frac{e^{2}_{h}}{6\pi^2}\sim \bigg\{\begin{array}{ccc}
                                                                                                      {\rm few}\times 10^{-10}&{\rm for}&{\mathcal{V}}\sim 10^{12} \\
                                                                                                      {\rm few}\times 10^{-20}&{\rm for}&{\mathcal{V}}\sim 10^{27}
  
\label{hyper}                                                                                                  \end{array}.
\end{equation}

\begin{figure}[tbp] 
\centering
{
\includegraphics[width=13cm,angle=0]{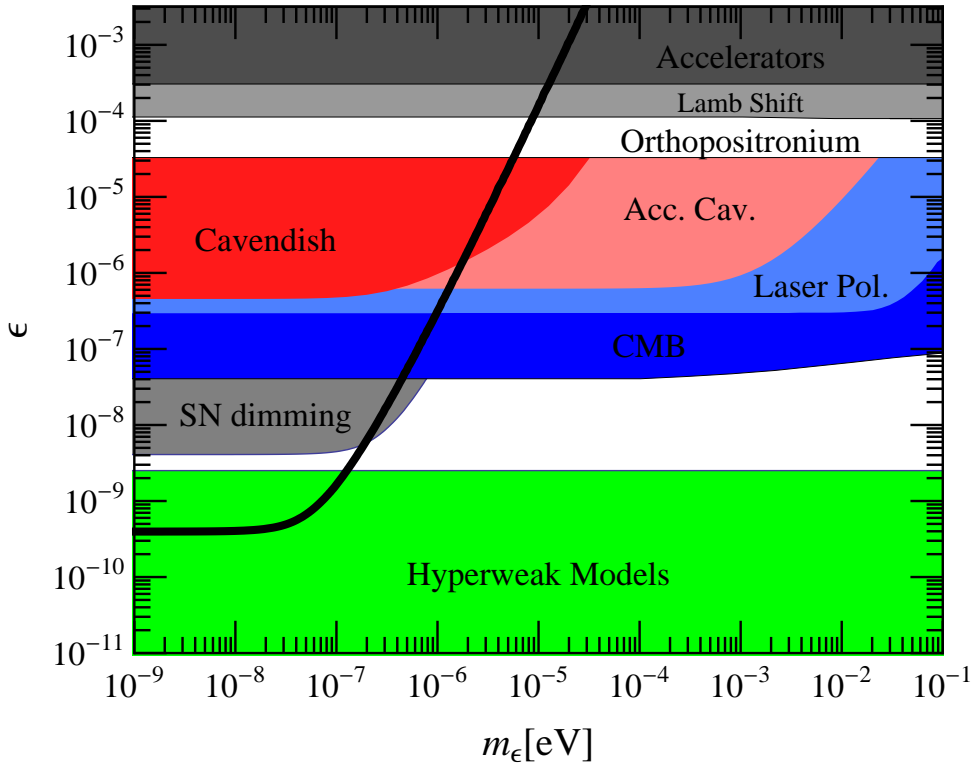}
}
\caption{Bounds on mini-charged particles for very weak hidden sector gauge couplings. They apply also to models with only mini-charged particles.
The solid black line shows the upper bound on the mini-charge obtained in this paper from the SZ effect.
The green area is a prediction in LARGE volume scenarios in string theory with a hyperweak U(1) and a string scale $M_{s}\lesssim10^{11}\,{\rm GeV}$.
For comparison, we have also included bounds arising from accelerators
\cite{Davidson:1993sj,Davidson:2000hf}, Lamb shift
\cite{Gluck:2007ia}, positronium decays \cite{Badertscher:2006fm}, tests
of Coulomb's Law \cite{Jaeckel:2009dh}, accelerator cavities \cite{Gies:2006hv}, laser polarization experiments
\cite{Ahlers:2007qf}, the CMB~\cite{Melchiorri:2007sq}
and supernova dimming~\cite{Ahlers:2009kh}.
All these bounds arise from physics occurring in low density/temperature regions.}
\label{fig:summary}
\end{figure}

Comparing with Fig.~\ref{fig:caca} we see that our bound probes the region of interest for the scenario with a string scale $M_{s}\sim 10^{11}\,{\rm GeV}$.
We have summarized this in Fig.~\ref{fig:summary} where we have also
included bounds from laboratory searches and low density/temperature
bounds from cosmology.  The solid line shows the edge of the excluded
region for models with only mini-charged particles. If the strength of
the magnetic field and the plasma frequency are assumed to be the same in each magnetic domain
in the cluster, for models with
hidden photons and hyperweak hidden gauge couplings satisfying (\ref{hyper}) there are `holes'
in the excluded region.  However if small fluctuations in the
magnetic field strength and plasma frequency are allowed, and such
fluctuations would naturally be expected to occur in the galaxy
cluster, these holes are closed and the entire region above the solid
line is excluded.

\section{The ISW effect: A future test}
\label{sec:ISW}
    On the largest scales on the sky ($\theta>1^{\circ}$)  the dominant source of secondary
    anisotropies of the CMB is not the SZ effect but the Integrated
    Sachs-Wolfe effect (ISW) \cite{Sachs:1967er}.  The ISW effect occurs when
    gravitational potentials evolve with time, as then the blue-shifting
    of a photon falling into the gravitational well is not exactly
    canceled by the red-shifting of the photon climbing out of the
    potential well, and there is a net effect on the energy of a
    photon,
\begin{equation}
\frac{\Delta T}{T}\approx-2 \int \dot{\Phi} d\tau,
\end{equation}
where $\Phi$ is the gravitational potential along a line of sight, and  a dot denotes differentiation with
respect to conformal time, $\tau$.

If the universe is  close to being flat most of
the late time ISW effect is caused by dark energy, and observations of
the
ISW effect have been used to probe its properties.  However measuring the ISW effect is not easy as the signal
is a fraction of the size of the primordial CMB anisotropies.  It can
be extracted, however, by looking for correlations between the CMB
temperature fluctuations and tracers of the density of matter.  These
correlations have been detected using a variety of density probes and
over a wide range of the electromagnetic spectrum \cite{Giannantonio:2008zi}.

The ISW effect dominates on the very largest scales, those of
superclusters of galaxies which can stretch over distances of tens of mega-parsecs.
We would expect mini-charged particles to be constrained by
observations of the ISW effect
only if there exist magnetic fields in galaxy superclusters.    So far no
detailed study has been done of the magnetic fields on such scales, however there are some hints that magnetic fields exist in these
environments from observations of the diffuse radio emission from
large scale networks of galaxies \cite{Bagchi:2002vf}.  Also if the magnetic fields
in galaxies are produced during structure formation (as opposed to
having their origin in a primordial magnetic field) then simulations
show that magnetic fields should indeed exist in superclusters of
galaxies (e.g. \cite{Ryu:1998up}).

It is to be hoped that as we continue to learn more abut the magnetic
fields in super clusters of galaxies we will be able to use
measurements of the ISW effect to constrain the properties of
mini-charged particles.  The magnetic fields in these structures are
expected, from simulations, to be of the same order of magnitude as
those in galaxy clusters and so we
expect to be able to significantly improve our bounds, both
because the temperature fluctuations in the CMB due to the ISW effect
are much smaller $\Delta T/T \sim 10^{-6}$ than those of the SZ
effect, and also because the distances traveled through the magnetic
fields of superclusters of galaxies are greater than the equivalent
distances used in the SZ calculation.

\section{Conclusions}
\label{sec:conc}
Mini-charged particles arise in a variety of extensions to the
Standard Model, most naturally in models which also contain  hidden photons.  In such models the
propagation of photons in a transverse magnetic field is affected by the real and
virtual production of MCPs.
MCPs can be constrained by a wide variety of terrestrial,
astrophysical and cosmological experiments.  However some of these
constraints, those  coming from processes occurring  in dense environments such as the
interior of stars, do not constrain all available MCP models. Therefore it is necessary to have
alternative probes of this region of parameter space which require only
physics in low density/temperature environments.  Such constraints are particularly
relevant for upcoming laboratory searches for MCPs.

In this article we have developed such a  new test for MCPs from
observations of the Sunyaev-Zel'dovich effect.  As photons from the
CMB pass through the magnetic fields of galaxy clusters some photons
are lost due to the production of MCPs and this decrement in the
photon flux  appears as an additional contribution to the SZ
effect.  Insisting that this flux decrement is not larger than the
observed SZ effect constrains new  regions of
the MCP parameter space.  Our bounds are most constraining for models
of MCPs without hidden sector photons, and for models with small
hidden sector gauge coupling which are well motivated in the LARGE
volume string scenario.

If, in the future, new observations lead us to a better understanding
of the magnetic fields in the largest scale structures in the
universe, a similar test could be made with observations of the ISW
effect, which  have
the prospect to be even more constraining for MCP models.

\section*{Acknowledgments}
We would like to thank Mark Goodsell for valuable discussions and
collaboration on related subjects.

\section*{References}
\bibliography{MCP}
\end{document}